\documentclass[prl, twocolumn, superscriptaddress]{revtex4}
\usepackage{amsmath}
\usepackage{graphicx}
\usepackage{amsfonts}
\usepackage{amssymb}

\begin{document}

\title{Single-particle and collective mode couplings associated with
1- and 2-directional electronic ordering in metallic RTe$_3$ (R =
Ho, Dy, Tb)}

\author{R. V. Yusupov}
\altaffiliation[Permanent address: ]{Kazan State University,
Kremlevskaya 18, 420008 Kazan, Russia} \email[E-mail:
]{Roman.Yusupov@ijs.si}  \affiliation{Complex Matter Dept., Jozef
Stefan Institute, Jamova 39, 1000 Ljubljana, Slovenia}

\author{T. Mertelj}
\affiliation{Complex Matter Dept., Jozef Stefan Institute, Jamova 39, 1000 Ljubljana,
Slovenia}

\author{J.-H. Chu}
\affiliation{Geballe Laboratory for Advanced Materials and
Department of Applied Physics, Stanford University, Stanford,
California 94305, USA}

\author{I. R. Fisher}
\affiliation{Geballe Laboratory for Advanced Materials and
Department of Applied Physics, Stanford University, Stanford,
California 94305, USA}

\author{D. Mihailovic}
\affiliation{Complex Matter Dept., Jozef Stefan Institute, Jamova 39, 1000 Ljubljana,
Slovenia}

\begin{abstract}
The coupling of phonons with collective modes and single-particle
gap excitations associated with one (1d) and two-directional (2d)
electronically-driven charge-density wave (CDW) ordering in metallic
RTe$_{3}$ is investigated as a function of rare-earth ion chemical
pressure (R=Tb, Dy, Ho) using femtosecond pump-probe spectroscopy.
From the $T$-dependence of the CDW gap $\Delta_{CDW}$ and the
amplitude mode (AM) we find that while the transition to a 1d-CDW
ordered state at $T_{c1}$ initially proceeds in an exemplary
mean-field (MF) -like fashion, below $T_{c1}$, $\Delta_{CDW}$ is
depressed and departs from the MF behavior. The effect is apparently
triggered by resonant mode-mixing of the amplitude mode (AM) with a
totally symmetric phonon at 1.75 THz.  At low temperatures, when the
state evolves into a 2d-CDW ordered state at $T_{c2}$ in the
DyTe$_3$ and HoTe$_3$, additional much weaker mode mixing is evident
but no soft mode is observed.
\end{abstract}

\maketitle

Ordered electronic states in condensed matter physics are of
fundamental importance as models for investigating the competition
between different ground states and collective behavior of quantum
systems. They are also of a practical interest because electronic
ordering gives rise to phenomena such as superconductivity and
colossal magnetoresistance which are all macroscopic manifestations
of underlying quantum phenomena. One class of systems which has
received renewed attention recently are two-dimensional (2D) layered
metals with electronically-driven charge-density-wave (CDW)
instabilities, partly because of the possible role that electronic
ordering may play in high-temperature superconductivity in layered
cuprates and (more recently) iron pnictides \cite{Pnictides}. In
weakly interacting systems, the instability is predominantly driven
by a Fermi surface (FS) nesting, where a single wavevector $q_N$
connects multiple points along the FS, giving rise to an enhanced
generalized susceptibility at this wavevector, which in turn leads
to the formation of an electronic ordered CDW state which reduces
the energy of the system. How this behavior evolves in strongly
coupled systems such as cuprates is still the subject of intense
investigations and is not clear.

\begin{figure}
\begin{center}
\includegraphics[angle=270,width=8cm]{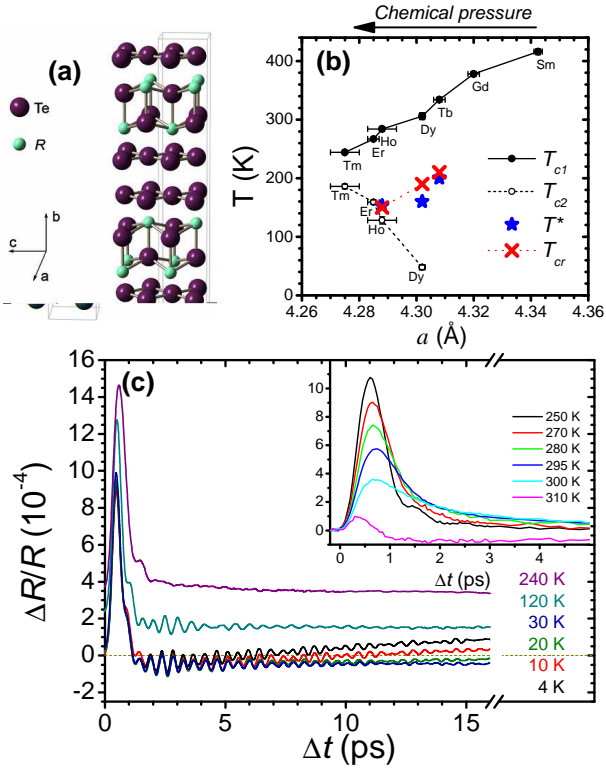}
\end{center}
\caption{(Color online). (a) The structure of RTe$_3$. (b) The phase
diagram of RTe$_3$ showing the critical temperatures $T_{c1}$ and
$T_{c2}$ for the 1d and 2d transitions respectively, as a function
of lattice constant $a$. Chemical pressure increases with decreasing
ionic radius of R (and $a$). (c) The raw data of the reflectivity
$\Delta R/R$ as a function of time delay $\Delta t$ of DyTe$_3$ at
different temperatures. The oscillations are the coherently excited
phonons and amplitude mode. The transient observed at short times is
the single particle response. The data for the other members of the
series are qualitatively the same.}\label{Fig1}
\end{figure}

Layered rare-earth tri-telluride compounds (RTe$_3$, where R is a
rare-earth ion) shown schematically in Fig.~\ref{Fig1}(a) are
interesting newly discovered examples of very weakly coupled
electronically-driven \cite{DIMASI:1994p2472} CDW systems whose
properties can be tuned by chemical pressure. Initial electron
diffraction \cite{DIMASI:1994p2472} and subsequent high resolution
x-ray diffraction \cite{Malliakas05, Malliakas06, Condron:2006p2469}
studies have revealed the ubiquitous presence of a weak lattice
modulation at a primary modulation vector $q_{CDW}$, where FS
nesting leads to the formation of a one-directional (1d)
incommensurate CDW. More recently, ARPES confirmed the existence of
imperfect FS nesting and revealed that as a result of the CDW, a gap
forms in the Fermi surface in the $\Gamma-Z$ direction
\cite{Gweon:1998p2468,Brouet:2008p2884} of the Brillouin zone.
Quantum oscillations from the reconstructed FS have been observed in
LaTe$_{3}$\cite{Borzi:2008p2880}. Recent theoretical work suggests
that there may be a competition between 1d ``stripe order" and 2d
``checkerboard order", which is finely tuned by the strength of the
electron-phonon coupling $\lambda$. Indeed, X-ray diffraction
\cite{Condron:2006p2469} and STM data \cite{Fang:2007p2466} on the
heavier rare-earth members of the series confirm the presence of a
``rectangular" 2d-ordered CDW state at low temperatures (at
$T_{c2}\simeq126$ K in HoTe$_3$ and 49 K in DyTe$_3$). In TbTe$_3$
there are also indications of possible 2d-ordering from
scanning-tunneling microscope studies at 6 K \cite{Fang:2007p2466},
but no long range ordering was seen by X-ray diffraction. In some
layered chalcogenides, notably in NbSe$_2$ pressure leads to
suppression of the CDW and emergence of superconductivity. Studies
as a function of chemical pressure in tri-tellurides have shown that
the 1d-CDW transition temperature $T_{c1}$ decreases with decreasing
R radius, but at the same time, another, coexisting 2d-ordered state
appears in the heavier R members of the series, whose critical
transition temperature $T_{c2}$ increases with increasing pressure
(see Fig.~\ref{Fig1}(b)). So far no superconductivity was
discovered, but all compounds show metallic resistivity down to the
lowest temperatures, in spite of CDW gaps on the FS. For most of the
series the local magnetic moments associated with rare earth ions
order antiferromagnetically at temperatures below the CDW $T_c$'s
\cite{Ru:2008AF}.

Femtosecond pump-probe spectroscopy (FPPS) has recently been shown
to be eminently suitable to the study of quasiparticle (QP) and
collective excitations of electronically ordered systems
\cite{YBCO,KMO,LSCO,2DCDW}. It allows the measurement of
low-frequency modes with very high resolution which are inaccessible
to Raman spectroscopy, as well as a direct measurement of the QP
recombination kinetics across the CDW or superconducting gap. In
this paper we use FPPS to investigate the evolution of the CDW gap
and the coupling of the amplitude mode and single particle
excitations with phonons in the 1d- and 2d-ordered states of three
tri-tellurides for the first time: HoTe$_3$ and  DyTe$_3$, which
exhibit two CDW transitions, and TbTe$_3$, for which there is only
one transition.

\begin{figure}
\begin{center}
\includegraphics[angle=270,width=6cm]{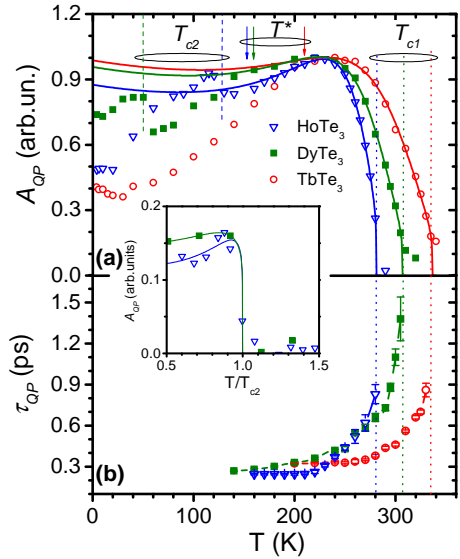}
\end{center}
\caption{(Color online). (a) The amplitude of the QP response as a
function of temperature for TbTe$_3$, DyTe$_3$ and HoTe$_3$. The
solid lines are the fits to the data using the predicted
single-particle response for a BCS-like gap function under
bottleneck relaxation conditions given by Eq.\ref{Eq1}. (b) The
relaxation time of the single particle response shows a divergence
as $T_{c1}$ is approached from below. }\label{Fig2}
\end{figure}

FPPS involves the measurement of the transient reflectivity response
after excitation by ultrashort (50 fs) laser pulses
\cite{KMO,2DCDW}. The laser pulses excite electron-hole pairs which
relax to states near the Fermi level in $< $50 fs by avalanche QP
multiplication. When there is a gap for electronic excitations at
low energy, such as is a CDW or superconducting gap, a relaxation
bottleneck may form, and a non-equilibrium population of the QPs at
the gap edge, which can be probed by excited state absorption with a
probe laser pulse, thus effectively measuring QP density in real
time. This QP density is usually assumed to be directly proportional
to the transient change of reflectivity $\Delta R/R$ which allows us
to directly observe the presence of a CDW gap, the QP relaxation
time $\tau_{QP}$, and their evolution with temperature.  In addition
to the QP excitations, with FPPS we also observe collective
excitations such as phonons and the AM \cite{KMO}, typically with
very high resolution and low noise. In our experiments the pump and
probe wavelengths were 400 and 800 nm, respectively. The pump
fluence was $\sim 20$ $\mu$J/cm$^2$, and laser heating was checked
to be minimal. The crystals used in this study were prepared by slow
cooling a binary melt, as described previously \cite{Ru:2006p2467}.
Clean surfaces oriented perpendicular to the $b$ axis of the crystal
were exposed by cleavage prior to the measurement.

The raw data on the transient reflectivity is shown in
Fig.~\ref{Fig1}(c)  for DyTe$_3$ (data on all three compounds are
qualitatively similar). The QP response gives rise to the short
transient, while the oscillations are from the coherently excited
phonons and the AM. The QP data are analyzed in the following way:
the maximum value of $\Delta R/R(t)$ was used as a measure of QP
density ($A_{QP}$), while the lifetime was obtained by fitting the
falling slope of the fast transient with $\Delta R/R(t) = A \exp
[-t/\tau_{QP}]$.  The fast transient is then subtracted from the
data to obtain the oscillatory responses, which are then analyzed
separately.

The amplitude of the QP response $A_{QP}$ and the lifetime
$\tau_{QP}$ for all three compounds is shown in Fig.~\ref{Fig2} as a
function of $T$. Their  temperature dependence is very similar near
$T_{c1}$. Fits to the data using a  theoretical model \cite{YBCO}
\begin{equation}\label{Eq1}
    A_{QP} \propto n_{pe}^* = \frac{\varepsilon_I/(\Delta(T) + k_B T/2)}
    {1 + \gamma \sqrt{\frac{2 k_B T}{\pi \Delta(T)}}\exp(-\Delta(T)/k_B T)}
\end{equation}
for the mean-field (MF) QP response using a BCS-like gap are shown
by the solid lines, showing remarkable agreement with the data near
$T_{c1}$.  The gap values for the Ho, Dy and Tb tellurides obtained
from the fit are $\Delta(0)_{1D}= 118(2), 123(3)$ and 125(6) meV,
respectively, in good agreement with previous optical measurements
\cite{Sacchetti:2006p2475} and somewhat less than the maximum gap
obtained in ARPES \cite{Brouet:2008p2884}, which we attribute to the
fact that optical measurements in general perform an average over
$\vec{q}$. The values of $\gamma$ were 20(5) for all traces. The
relaxation time is theoretically related to the gap as $\tau_{QP}
\propto 1/\Delta_{1d}(T)$ near $T_c$ \cite{YBCO}, so the divergence
of $\tau_{QP}$  for all three compounds is further remarkable
indication of MF behavior, where $\Delta_{1d}(T) \rightarrow 0$ as
$T \rightarrow T_{c1}$.

In contrast to the behavior near T$_{c1}$, below $\thickapprox 200$
K we see a systematic departure from the predicted MF behavior in
all three compounds. The order parameter in TbTe$_3$ observed by
x-ray diffraction shows a similar departure from MF behavior in the
same temperature range \cite{Condron:2006p2469}. In our case, two
effects are visible: a drop in the amplitude below the MF
prediction, and a small but systematic gap-like feature which
coincides with $T_{c2}$ in DyTe$_3$ and HoTe$_3$.  The insert to
Figure 2 shows that the gap-like anomaly at $T_{c2}$ can be fit to
the appearance of an additional CDW gap opening for both Ho and Dy
tellurides. The fit gives values of the gap at: $\Delta(0)_{2D} =
40(4)$ meV  and $\Delta(0)_{2D} = 13.5(1.5)$ meV for HoTe$_3$ and
DyTe$_3$ respectively with $\gamma = 3$. No such gap opening is
unambiguously seen in TbTe$_3$, where the existence of a second
transition is uncertain \cite{Condron:2006p2469}. Turning to the
departure from MF behaviour, we have plotted the point where the
amplitude $A_{QP}$ deviates from the MF prediction on the phase
diagram in Fig.~\ref{Fig1}. The deviation is systematic in all three
compounds and occurs in the temperature range 140 K $< T^{*} < 200$
K.

\begin{figure}
\begin{center}
\includegraphics[angle=270,width=8cm]{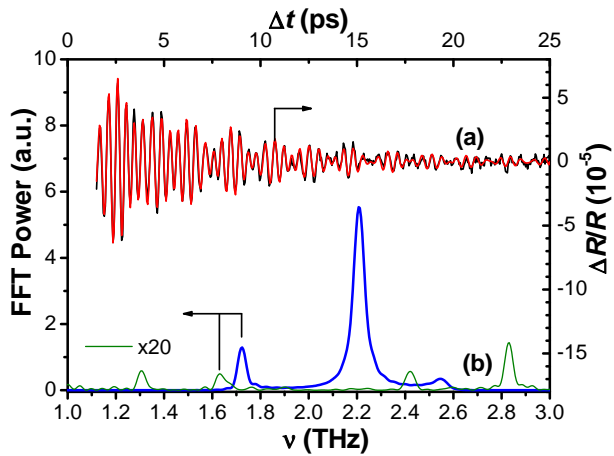}
\end{center}
\caption{(Color online). (a) The oscillatory response in DyTe$_3$ at
3.8 K (black) with a fit using three damped oscillators (red). (The
relaxation component is subtracted.) (b) The fast Fourier transform
of the fitted trace shown in (a) (thick), and the FFT of the
remaining modes (thin line).} \label{Fig3}
\end{figure}

\begin{figure}
\begin{center}
\includegraphics[width=7cm]{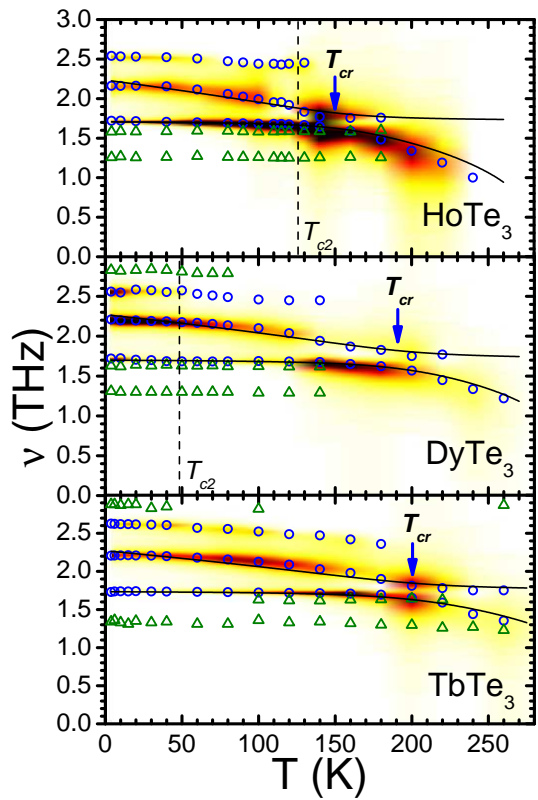}
\end{center}
\caption{(Color online). A phonon intensity map as a function of $T$
and $\nu$ for HoTe$_3$, DyTe$_3$ and TbTe$_3$ showing all the
observed oscillators. Mode mixing is evident between the AM (whose
frequency is 2.2THz at low $T$), the 1.75 THz mode at intermediate
temperatures, and with the 2.6 THz phonon at low temperatures. The
latter is particularly evident in DyTe$_{3}$. The lines are fits to
the data using a simple level-crossing model calculation (see
text).}\label{Fig4}
\end{figure}

Next, let us turn our attention to the AM and the phonons. To
minimize problems with fast Fourier transform (FFT) artifacts in the
data analysis, we fit the coherent phonon oscillations in two
stages. First we use a set of three oscillators which are fit in
time-domain, subtract them from the data, and then the residual
time-trace containing weak oscillations is analyzed separately. As a
typical example, the raw time-domain data and FFTs of the fitted
time-trace and residual time-trace are shown in Fig.~\ref{Fig3} for
DyTe$_{3}$ at $T=3.8$ K. In Fig.~\ref{Fig4} we show the evolution of
the observed modes as a function of temperature. Following the mode
assignments of Lavagnini \textit{et al.} \cite{Raman}, in the
distorted phase there are 56 $A_{1}$ and 28 $B_{1}$-symmetry modes,
all of which we can observe in principle. However, similarly as in
Raman, many may be overlapping and we observe only a few. Many of
the phonon modes are observed to shift in frequency at low
temperatures. The color map (Fig.~\ref{Fig4}) shows also the
linewidths and intensities of the most intense modes. At low $T$,
the modes are very sharp and quite intense, but near $T_{c1}$, all
modes virtually disappear.  A mode at 2.2 THz - which we attribute
to the AM of the CDW - is seen to soften significantly on
approaching $T_{c1}$ from below.  Resonant coupling of the AM with
the mode at 1.75 THz is clearly seen in all three compounds. In
addition, the mode near 2.6 THz is observed to show a frequency
renormalization, which may be associated with the transition to
bidirectional order at $T_{c2}$ in Ho and Dy tellurides. In
TbTe$_{3}$ we see no further effect at low $T$, consistent with the
absence of low-temperature CDW ordering in this compound.

In Fig.~\ref{Fig4}, we also show a model fit which describes the
crossing of the AM with the 1.75 THz phonon. We use a simple model
of two interacting vibrational modes, where the AM has a  MF-like
$T$-dependence $\nu_{AM} = \nu_{0}*(1-T/T_c)^{\beta}$ and $\nu_{p} =
1.75$ THz, with an interaction term (off-diagonal matrix element)
$\delta$. The model is solved analytically, and the fits to the data
in Fig.~\ref{Fig4} give the same rather large value for the
AM-phonon interaction in all three compounds, $\delta = 0.105 \pm
0.011$ THz $(3.5 \pm 0.5$ cm$^{-1}$), and $\beta = 0.30(3)$.

Remarkably, the temperature $T_{cr}$ at which the AM frequency
crosses the 1.75 THz phonon is very close to $T^{*}$ for all three
tellurides (compare Figs.~\ref{Fig2} and \ref{Fig4}). Furthermore,
we see that the 2.6 THz phonon for all three compounds appears just
below $T_{cr}$. From this systematic behavior we conclude that the
anomalies in the QP density are related to the strong mixing of the
AM with the 1.75 THz phonon. In the absence of a phonon mode
assignment for the 1.75 THz mode, we cannot analyze the ionic
displacements quantitatively, but we can qualitatively understand
the observed effect as follows. Since the 1.75 THz phonon is clearly
coupled to the AM, by symmetry this mode is also directly coupled to
the charge excitations of the charge density wave. The strong mixing
of the two modes will thus disturb the AM Te ion displacements by
introducing other displacements corresponding to the 1.75 THz
phonon, which will in turn have the effect of reducing the AM
amplitude, and depress the CDW gap $\Delta_{CDW}$. In other words,
the CDW gap is renormalized by the 1.75 THz mode once it mixes with
the AM. The appearance of the 2.6 THz phonon can then be understood
to arise from symmetry breaking below $T^*$.


Summarizing the interplay between the different signatures of
low-temperature ordering we see that the onsets of 1d- and
2d-ordering are clearly seen in the QP response. Judging from the
fits to the data, the increase in amplitude which coincides with
$T_{c2}$ is most likely associated with the opening of a second gap
which accompanies the low-temperature transition to 2-directional
ordering. High resolution XRD \cite{Condron:2006p2469} and ARPES
\cite{Moore:unpub} in ErTe$_{3}$ indeed suggest the opening of an
additional gap along the $\Gamma-X$ direction. In contrast, in the
phonon spectrum, an AM is associated only with the 1d transition at
$T_{c1}$, and in spite of high resolution and excellent
signal-to-noise in our data, we do $not$ observe any soft mode which
can be associated with $T_{c2}$.

However, the QP response at low $T$ in the three compounds
investigated here cannot be simply attributed to just the opening of
a gaps at $T_{c1}$ and $T_{c2}$. The clear correlation between the
temperature $T^{*}$ where the QP response departs from MF behaviour,
the mode crossing temperature $T_{cr}$ and the appearance of the 2.6
THz phonon in all three compounds is unambiguous indication of
systematic disruption of 1d-CDW order resulting from coupling of a
phonon at 1.75 THz to the CDW, and the drop of amplitude of the QP
response well below $T^{*}$ seen in Fig.~\ref{Fig2} implies a
renormalization of the gap. The increasing $T^{*}$ (or $T_{cr}$)
with decreasing rare earth pressure is a direct consequence of the
increasing $T_{c1}$, which shifts the crossing temperature $T_{cr}$
of the AM with the 1.75 THz phonon. Finally, we note that the strong
soft mode behavior of the AM is clear indication that the transition
at $T_{c1}$ in DyTe$_3$  and HoTe$_3$ is of second order. From the
fits to the $T$-dependence of the QP response it would appear that
the low-temperature phase transition is also second order, but since
no soft mode is observed associated with this transition, this
cannot be claimed with certainty. Nevertheless, none of the observed
QP responses appear to show any temperature-hysteresis associated
with either $T_{c1}$ or $T_{c2}$, apparently confirming this
assignment. Absence of an additional AM developing below $T_{c2}$ is
surprising by itself as these excitations have been observed in all
systems studied so far, the AM is totally-symmetric and thus should
be present in the coherent phonon spectra.

\begin{acknowledgments}
Studies at Jozef Stefan Institute are supported within the FP6,
project NMP4-CT-2005-517039 (CoMePhS). The work at Stanford
University is supported by the DOE, Office of Basic Energy Sciences,
under contract No. DE-AC02-76SF00515.
\end{acknowledgments}

\end{document}